\documentclass{ifacconf}

\usepackage{graphicx}      
\usepackage{natbib}        
\usepackage{times} 
\usepackage{graphicx}  
\usepackage{caption}
\usepackage{subcaption}

\usepackage{array}
\usepackage{tikz}
\usepackage{latexsym,theorem}
\usepackage{amsmath,amsfonts,amssymb,amsbsy}
\usepackage{verbatim}
\usepackage{mathrsfs}

\usepackage{epstopdf}
\usepackage{color}
\usepackage{algorithm}
\usepackage{mathtools}
\usepackage[noend]{algpseudocode}
\DeclarePairedDelimiter{\ceil}{\lceil}{\rceil}

\newcommand{\Uset}{\mathbb{U}}

\newcommand{\Xset}{\mathbb{X}}
\newcommand{\Fset}{\mathbb{F}}

\newcommand{\Pset}{\mathbb{P}}

\newcommand{\cS}{\mathcal{S}}
\newcommand{\cU}{\mathcal{U}}

\newcommand{\cK}{\mathcal{K}}

{\theorembodyfont{\slshape}\newtheorem{theorem}{Theorem}[section]}
{\theorembodyfont{\slshape}}
{\theorembodyfont{\slshape}}
{\theorembodyfont{\slshape}}
{\theorembodyfont{\upshape}}
{\theorembodyfont{\upshape}\newtheorem{problem}[theorem]{Problem}}
{\theorembodyfont{\upshape}\newtheorem{remark}[theorem]{Remark}}
{\theorembodyfont{\upshape}}
{\theorembodyfont{\upshape}}
{\theorembodyfont{\upshape}}

\begin{document}
\begin{frontmatter}

\title{Towards Parallelizable Sampling--based \\ Nonlinear Model Predictive Control}


\author[First]{R.V. Bobiti}
\author[First]{M. Lazar}

\address[First]{Department of Electrical Engineering, Eindhoven University of Technology, The Netherlands (e--mails: r.v.bobiti@tue.nl, m.lazar@tue.nl)}

\begin{abstract}                
This paper proposes a new sampling--based nonlinear model predictive control (MPC) algorithm, with a bound on complexity quadratic in the prediction horizon $N$ and linear in the number of samples. The idea of the proposed algorithm is to use the sequence of predicted inputs from the previous time step as a warm start, and to iteratively update this sequence by changing its elements one by one, starting from the last predicted input and ending with the first predicted input. This strategy, which resembles the dynamic programming principle, allows for parallelization up to a certain level and yields a suboptimal nonlinear MPC algorithm with guaranteed recursive feasibility, stability and improved cost function at every iteration, which is suitable for real--time implementation. The complexity of the algorithm per each time step in the prediction horizon depends only on the horizon, the number of samples and parallel threads, and it is independent of the measured system state. Comparisons with the \emph{fmincon} nonlinear optimization solver on benchmark examples indicate that as the simulation time progresses, the proposed algorithm converges rapidly to the ``optimal'' solution, even when using a small number of samples.







\end{abstract}


\end{frontmatter}

\section{Introduction}

A vast literature on nonlinear model predictive control (NMPC) has proven both the theoretical \citep{grune2011nonlinear}, as well as the practical advantage \citep{magni2009nonlinear} of this method in treating optimally the control of multi--variable nonlinear systems subject to constraints on state and inputs. Common research interests include methods for reducing the complexity of the NMPC algorithms, to make them applicable on devices of low memory (ASIC, FPGA), such as explicit NMPC (ENMPC), see, e.g., \citep{johansen2004approximate}. Much work has been done on treating other limiting factors, such as the requirement of NMPC of solving an optimization problem online. This is not well achieved by common optimization tools, which have no specific termination time, especially due to non--convexity which involves multiple local--minima. Therefore, real time requirements are not met, which limits the industrial impact of NMPC. To alleviate these concerns, the literature has proposed multiple solutions, such as approximate dynamic programming (DP) \citep{bertsekas1995dynamic}, suboptimal MPC \citep{scokaert1999suboptimal}, approximative DP and suboptimal control via rollout algorithms \citep{bertsekas2005dynamic}, NMPC based on system approximation by neural models \citep{lawrynczuk2009computationally}.

An alternative strategy in NMPC is to draw samples from either the state or input space, to design computationally feasible NMPC methods, see for example, \citep{piovesan2009randomized} which proposes a randomized approach to sampling the space of predicted input sequences. More recently, in \citep{chakrabarty2016}, an ENMPC method was proposed based on sampling of the state space for continuously differentiable nonlinear systems. The method therein solves optimization problems offline to find optimal control sequences, which are used to construct the ENMPC strategy. While there are still concerns in ENMPC related to robustness, feasibility of the offline optimization and finding the neighbors in the sampled grid for the off--grid states, ENMPC, when successful, reduces significantly the computational load of MPC at the expense of an acceptable cost degradation.

Input and state space sampling--methods for solving NMPC via approximate DP have also been proposed, see \citep{bertsekas1995dynamic}, though they inherit the dimensionality issues of DP \citep{lee2004approximate}.

Another relevant sampling--based strategy, the so--called sampling based MPC (SBMPC), was proposed in \citep{dunlap2010nonlinear}. The method therein is applicable to nonlinear systems in general, though, its performance is dependent on a user--specified heuristic. A more ample discussion on sampling--based DP and SBMPC, in the light of the method proposed in this paper is reported in Section~\ref{existing}.

A common problem of sampling--guided methods for NMPC is the sampling strategy. 
For example, with each input sample, a tree is expanded. After the tree is built, the path of least cost in the tree is selected from the initial state to the desired state. If the sampling is performed over the input space, and each sample is connected to all the samples in the input space for the next time step in the control horizon, then the tree growth is exponential with the horizon. Alternatively, as in randomized MPC \citep{piovesan2009randomized}, sampling randomly in the input space, of dimension $m$, augmented to the horizon of dimension $N$ requires a large number of samples, in an $mN$ dimensional space, to achieve a significant accuracy.

In this paper we adopt a suboptimal formulation of NMPC, as originally proposed in \citep{scokaert1999suboptimal}, where it was shown that feasibility of a solution implies stability under suitable conditions. This, together with the fact that suboptimal NMPC has the same inherent robustness properties as optimal NMPC, see \citep{pannocchia2011conditions} and \citep{lazar2009predictive}, suggest that suboptimal NMPC is a viable and in fact the best one can hope for when a sampling--guided MPC strategy is undertaken for the control of nonlinear systems. Furthermore, we aim at a sampling method which provides a suboptimal solution that yields good control performance, has a reasonable computational complexity increase with the prediction horizon and allows for parallel implementation up to some level.

In this paper, the main idea for achieving this goal is to use the shifted sequence of predicted inputs from the previous time step as a warm start, and to iteratively update this sequence by changing its elements one by one, starting from the last predicted input and ending with the first predicted input. This strategy resembles the dynamic programming principle, especially the rollout algorithms, see \citep{bertsekas2005dynamic}, which improve a heuristic base policy for optimal control. Additionally, in this paper, we sample the original input space, which is typically represented by a proper set $\Uset\subset\Rset^m$. Sampling allows for parallelization of the calculations performed for updating each of the elements of the predicted sequence of inputs and it enables limiting the computational time according to the requirements of the considered application. An upper--bound on the complexity of the overall algorithm is quadratic with the prediction horizon $N$ and linear with the number of samples in $\Uset$. This enables the usage of long prediction horizons or real--time implementation on inexpensive computing devices such as ASIC and FPGA. The suitability for real--time implementation is also enhanced by the fact that the algorithm can be stopped at any iteration performed within a sampling period, while the complexity of the calculations per iteration depends only on the horizon $N$, the number of samples and parallel threads, and it does not depend on the measured state of the system. Moreover, the updated predicted sequence of inputs obtained at any iteration will guarantee recursive feasibility, stability and an improved cost function under the same conditions as in suboptimal NMPC \citep{scokaert1999suboptimal}.

The remainder of this paper is organized as follows. In Section~\ref{preliminaries}, basic notation is introduced, together with the problem formulation and a discussion on the relation with the existing methods. Section~\ref{main_res} presents the main result as a prototype algorithm and its complexity analysis. In Section~\ref{examples}, three examples illustrate the potential of the developed method, and Section~\ref{conclusions} concludes the paper.




\section{Preliminaries}
\label{preliminaries}

\subsection{Notation}
\label{notation}

Let $\mathbb{R}$, $\mathbb{R}_+$, $\mathbb{Z}$ and $\mathbb{Z}_+$ denote the field of real numbers, the set of non--negative reals, the set of integers and the set of non--negative integers, respectively. For every $c\in\Rset$ and $\Pi\subseteq\Rset$, define $\Pi_{\geq c}:=\{k\in\Pi\mid k\geq c \}$ and similarly $\Pi_{\leq c}$. Let $int(\mathbb{S})$ denote the interior of a set $\mathbb{S}$. Let $\mathbb{S}^h:=\mathbb{S}\times\ldots\times\mathbb{S}$ for any $h\in\mathbb{Z}_{\geq1}$ denote the $h$--times Cartesian--product of $\mathbb{S}\subseteq\Rset^n$.
A set $\cS\subset\Rset^n$ is called proper if it is non--empty, compact and $0\in int(\cS)$.

For a vector $x\in\Rset^n$, the symbol $\|x\|$ is used to denote an arbitrary $p$--norm; it will be made clear when a specific norm is considered. For a vector $x\in\Rset^n$, define $|x|:=[|x_1|\ldots|x_n|]^T$. Also for a vector $x\in\Rset^n$, by $\max|x|$ we denote $\max \{|x_1|,\ldots,|x_n|\}$. For a scalar $x\in\Rset$, denote by $\ceil{x}$ the smallest integer number larger than $x$.

A function $\alpha:\mathbb{R}_+\rightarrow\mathbb{R}_+$ is said to belong to class $\cK$, i.e., $\alpha\in\cK$, if it is continuous, strictly increasing and $\alpha(0)=0$. Furthermore, $\alpha\in\cK_\infty$ if $\alpha\in\cK$ and $\lim_{s\rightarrow\infty}\alpha(s)=\infty$.

\subsection{Suboptimal MPC problem formulation}
\label{mpc}

Let us consider the discrete--time system described by
\begin{align}\label{eq:system}
x_{k+1}=f(x_k, u_k),
\end{align}
where $x_k\in\Rset^n$ is the state and $u_k\in\Rset^m$ is the control vector at discrete--time $k\in\Zset_+$. We assume that the map $f:\Rset^n\times\Rset^m\rightarrow \Rset^n$ satisfies $f(0, 0)=0$, which is, the origin is an equilibrium point for system \eqref{eq:system}.

The goal of MPC is to regulate the state to the origin while satisfying control and state constraints, i.e., $u_k\in\Uset\subset\Rset^m$ and $x_k\in\Xset\subset\Rset^n$ for all $k\in\Zset_+$, where $\Uset$ and $\Xset$ are proper sets. MPC relies on a receding--horizon control law in order to determine, for each $k$, a finite--sequence of control inputs
$$U(k)=\{u_{k|k}, u_{k+1|k}, \ldots, u_{k+N-1|k}\},$$
where $N$ is the control and prediction horizon, which are considered equal in this paper, for simplicity of exposition.
If the initial state is $x_{k|k}=x_{k}$ and the control sequence is $U(k)$, the solution of system \eqref{eq:system} in closed-loop with $U(k)$ at time $k+i$ is denoted by $\phi(x_{k|k}, U(k), i)$. The current control action $u_k$, is selected as the first control action in $U(k)$, i.e., $u_k=u_{k|k}$.

To achieve this, optimal MPC minimizes, at each discrete--time $k$, a cost function of the type
\begin{align}\label{eq:cost}
J(x_{k|k}, U(k))=V_f(x_{k+N|k})+\sum_{i=0}^{N-1}L(x_{k+i|k}, u_{k+i|k}),
\end{align}
where $J:\Rset^n\times\Rset^m\rightarrow\Rset_+$ is the total cost function,  $V_f:\Rset^n\rightarrow\Rset_+$ is a terminal cost and $L:\Rset^n\times\Rset^m\rightarrow\Rset_+$ is a stage cost. The minimization is performed with respect to $U(k)$ and it is subject to
\begin{align}\label{eq:eqconstr}
x_{k+j|k}=f(x_{k+j-1|k}, u_{k+j-1|k}), \quad \forall j\in\Zset_{[1, N]},
\end{align}
and to the state and input constraints.

When the dynamics $f$ is a nonlinear, possibly non-convex function, the optimization of the cost \eqref{eq:cost} cannot be guaranteed to converge to a global optimum, in general. Alternatively, suboptimal MPC was proposed as a viable alternative, see, e.g., \citep{scokaert1999suboptimal}, to deal with this inherent shortcoming of nonlinear global optimization.

Assume that the following constraints are required to hold at each iteration of the MPC problem:
\begin{equation}\label{eq:constraints}
x_{k+i|k}\in\Xset, u_{k+i|k}\in\Uset, \quad \forall i\in\Zset_{[0, N-1]},
\end{equation}
and
\begin{equation}\label{eq:termcon}
 x_{k+N|k}\in\Xset_T,
\end{equation}
where $\Xset_T\subseteq\Xset$ is a proper set which represents a terminal constraint. Moreover, define by $\cU(x_{k|k})$ the set of control sequences $U(k)$ which, applied on $x_{k|k}$, satisfy \eqref{eq:eqconstr}, \eqref{eq:constraints} and \eqref{eq:termcon}.

Suboptimal MPC relies on an initial feasible solution, a warm start sequence $U_{warm}(k)\in \cU(x_{k|k})$ at each step $k$, which is improved iteratively. The suboptimal MPC problem considered in this paper is formulated as follows:

\begin{problem}\label{prob}
For each $k\in\Zset_{\geq 1}$, given $U_{warm}(k)$ find a sequence $U(k)\in\cU(x_{k|k})$ such that
\begin{align}\label{eq:subMPC}
J(x_{k|k}, U_{warm}(k))> J(x_{k|k}, U(k)),
\end{align}
and the constraints \eqref{eq:eqconstr}, \eqref{eq:constraints} and \eqref{eq:termcon} are satisfied.
\end{problem}

Consider now a locally stabilizing control law $k_f:\Xset_T\rightarrow\Uset$. Assume that $\Xset_T$ is a sublevel set of $V_f$. For stability of the MPC closed--loop system it is also required that:
\begin{itemize}
\item $V_f(f(x, k_f(x)))+L(x, k_f(x))\leq V_f(x)$ for all $x\in\Xset_T$;
\item there exist $\alpha_1, \alpha_2\in\cK_{\infty}$ such that $\alpha_1(|x|)\leq V_f(x)\leq \alpha_2(|x|)$ for all $x\in\Xset_T$;
\item there exist $\alpha_3\in\cK_{\infty}$ such that $L(x, u)\geq \alpha_3(|x|)$ for all $(x, u)\in\Xset\times\Uset$;
\end{itemize}

\begin{remark}
The first property above listed implies that $\Xset_T$ is positively invariant for the system $x_{k+1}=f(x_k, k_f(x_k))$. The second and third properties can be satisfied if, for example $V_f$ and $L$ are positive definite quadratic functions.
\end{remark}
 These properties imply that the cost function $J(\cdot, \cdot)$ is a Lyapunov function, see \citep[Lemma 2.14]{mayne2009model}. As such, if $\Fset$ is the set of states in $\Xset$ for which there exists a control sequence $U(k)$ which satisfies the constraints \eqref{eq:eqconstr}, \eqref{eq:constraints} and \eqref{eq:termcon}, then the solution to Problem~\ref{prob} provides an asymptotically stabilizing controller with a region of attraction $\Fset$.

In \citep{mayne2009model}, an algorithm was proposed for suboptimal MPC with stability guarantees. Given the current state $x_{k|k}$ and the previous control sequence $U(k-1)$ as an input, the steps of the algorithm therein can be summarized as follows:
\begin{itemize}
\item If $x_{k|k}\notin\Xset_T$, use the warm start sequence:
\begin{align}
U_{warm}(k)= & \{u_{k|k-1}, u_{k+1|k-1}, \ldots, \nonumber \\
& u_{k+N-2|k-1}, k_f(x_{k+N-1|k-1})\}.
\end{align}
Solve iteratively Problem~\ref{prob} via optimization to improve $U_{warm}(k)$ with a $U(k)\in\cU(x_{k|k})$. Apply control $u_k=u_{k|k}$.
\item If $x_{k|k}\in\Xset_T$, set $u_{k|k}=k_f(x_{k|k})$, or, similarly to the previous case, use the warm start and solve an optimization algorithm iteratively to find an improved control sequence $U(k)\in\cU(x_{k|k})$.
\end{itemize}

Optimization solvers, in both optimal and suboptimal MPC, present difficulties in terms of parallelization and a priori known execution time independently of the current state $x_{k}$.
To circumvent these drawbacks and enable a computationally efficient and parallelizable nonlinear MPC algorithm, we propose a sampling based approach to solving Problem~\ref{prob}, as illustrated in Section~\ref{main_res}. To this end, we first review existing, similar approaches within nonlinear MPC.

\subsection{Existing approaches based on sampling}
\label{existing}

This section provides a brief in depth review of two main existing approaches for NMPC based on sampling, namely approximate DP and sampling--based MPC, which were also mentioned in the Introduction.

An approximate version of DP, as a tool for solving optimization problems, was proposed in \citep[Section 6.6.1]{bertsekas1995dynamic}. DP has been successfully applied for determining explicit solutions for linear MPC controllers, see, e.g., \citep{muiioz2004dynamic}. In nonlinear MPC, the state and input states are typically discretized to apply DP algorithms. The main idea is to discretize the state space with a finite grid and to express each state outside of the grid as an interpolation of nearby grid elements. The same interpolation law is applied to compute the cost of the current nongrid state as a function of the costs of the nearby grid states. As such, by discretizing both the state space for each time in the control horizon and the input space, a transition diagram is obtained which approximates the dynamics of the system in the continuous space. On this discrete transition system, DP is applied to determine the path with the smallest cost, which, for a given initial state $x_{k|k}$, provides the control sequence $U(k)$.

Solving MPC with DP via discretization suffers from the ``curse of dimensionality", due to sampling of both state and input spaces and the requirement for constructing the complete transition diagram, by evaluating the subsequent state and cost for each sampled state and all the samples in the input space.

An alternative to approximate DP, namely sampling based MPC, was developed within the area of Robotics, where typically optimization problems arising in control are non--convex, due to either kinematic constraints or constraints posed by obstacle avoidance.
Sampling--based motion planning such as rapidly-exploring random trees (RRTs) \citep{lavalle1998rapidly} or randomized A* algorithms \citep{likhachev2008r}, have been extensively used to construct trees which connect an initial state to a final state based on sampling states in the search space and searching for feasible inputs to connect these states. To approach issues related to the, possibly unfeasible, search for an input after sampling only in the state space, a method (SBMPC) was proposed in \citep{dunlap2010nonlinear}, which samples the input space at each sampling period and creates trees that contain feasible state trajectories. The optimal path to a goal in the state--space is then searched for within the tree using goal--directed search algorithms, such as LPA*. Such algorithms rely on computing a heuristic measure of the distance from the current sample to the goal. Selecting the heuristic is however, not an obvious task  for general nonlinear systems.

Therefore, a desirable feature of a sampling--based suboptimal NMPC algorithm is a non--exponential growth in the tree generated through sampling of the state or input space at each step in the horizon. Furthermore, it is also desirable to reduce the dependency of the algorithm on the non--obvious selection of a heuristic, which significantly impacts the performance of the sampling--based strategy. To circumvent these issues, an alternative suboptimal strategy for sampling the input space is proposed in the next section, based on sequentially updating a warm start feasible sequence of predicted inputs.


\section{Main result}
\label{main_res}
In this section we present the proposed sampling--based algorithm for solving suboptimal NMPC problems and we provide a detailed complexity analysis.
\subsection{Prototype algorithm}
\label{proto-alg}

Inspired by the suboptimal MPC and DP principles, we propose a sampling--based solution to Problem~\ref{prob}, which, by the mechanism involved in the iterative improvement of the initial feasible control sequence $U_{warm}(k)$, has a low increase with the control horizon of the computational complexity.

\begin{figure}[!htpb]
      \centering
      \includegraphics[width=1\columnwidth]{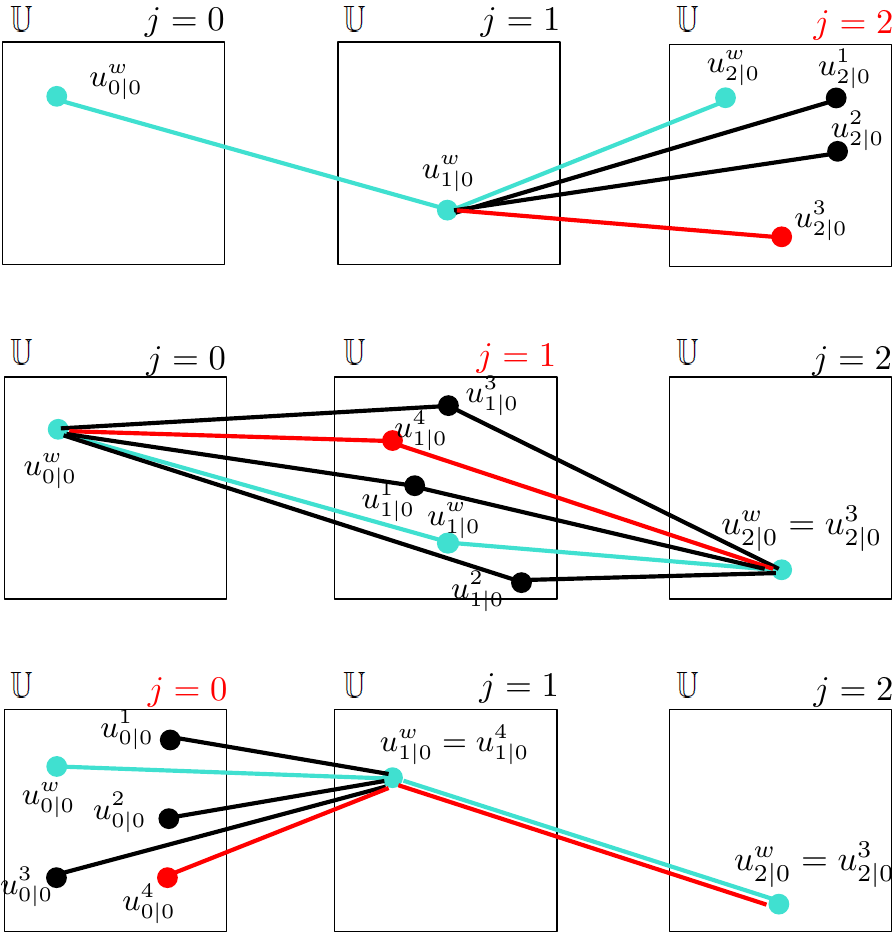}
      \caption{Sampling principle.}
      \label{sample}
\end{figure}

\algnewcommand{\And}{\textbf{and}}
\renewcommand{\algorithmicrequire}{\textbf{Input:}}
\renewcommand{\algorithmicensure}{\textbf{Output:}}
\begin{algorithm}
\caption{Sampling--based suboptimal NMPC algorithm.}\label{alg}
\begin{algorithmic}[1]
\Require $N$, $\{n_j\}_{j\in\Zset_{[0, N-1]}}$, $x_{k|k}$, $\Xset$, $\Uset$, $\Xset_T$, $J(\cdot, \cdot)$

$U_{warm}(k)=\{u_{k|k}^w, \ldots, u_{k+N-1|k}^w\}$
\Ensure $U(k)$, $u_k$
\Statex
\State $J_{sub}\leftarrow J(x_{k|k}, U_{warm}(k))$;
\ForAll{$j=N-1:-1:0$}
        \State Select $n_j$ samples $u_{j+k|k}^q\in\Uset$, $q\in\Zset_{[1,...,n_j]}$;
        \ForAll{$q=1:1:n_j$}
            \If{$j\geq 1$}
                \State $U(k)^{jq}=\{u_{k|k}^w, \ldots, u_{k+j-1|k}^w, u_{k+j|k}^q,$

                $ \quad\quad\quad\quad\quad\quad\quad u_{k+j+1|k}^w,\ldots, u_{k+N-1|k}^w\}$;
            \Else
                \State $U(k)^{jq}=\{u_{k|k}^q, u_{k+1|k}^w, \ldots, u_{k+N-1|k}^w\}$;
            \EndIf
            \If{$\phi(x_{k|k}, U(k)^{jq}, i)\in\Xset, \forall i\in\Zset_{[j+1, N-1]}$ and $\phi(x_{k|k}, U(k)^{jq}, N)\in\Xset_T$}
                \State $J_{new}\leftarrow J(x_{k|k}, U(k)^{jq})$;
                \If{$J_{new}<J_{sub}$}
                    \State $J_{sub}\leftarrow J_{new}$;
                    \State $U_{warm}(k)\leftarrow U(k)^{jq}$;
                \EndIf
            \EndIf
        \EndFor
\EndFor
\State $U(k)=U_{warm}(k)$, $u_k=u_{k|k}^w$;

\end{algorithmic}
\end{algorithm}

The principle behind the proposed sampling--based approach is illustrated in Fig.~\ref{sample} and formalized in Algorithm~\ref{alg}. In Fig.~\ref{sample}, the iterative improvement of an initial cost provided by an initial feasible control sequence $U_{warm}(k)$ is illustrated for the case when $N=3$. The algorithm keeps always $U_{warm}(k)$ as a reference sequence, and it covers the horizon in a backward fashion, in $N$ iterations. Starting with $j=N-1$, at each iteration step, $n_j$ samples $\{u_{k+j|k}^q\}_{q\in\Zset_{[1, n_j]}}$ are drawn from the input constraint set $\Uset$. For each sample, the reference sequence $U_{warm}(k)$ is modified in the $j^{th}$ location, and a new sequence, $U(k)^{jq}$ is obtained. If the following properties hold:
\begin{itemize}
\item $U(k)^{jq}$ is a feasible sequence, i.e., the constraints \eqref{eq:eqconstr}, \eqref{eq:constraints} and \eqref{eq:termcon} hold,
\item the new cost, $J_{new}=J(x_{k|k}, U(k)^{jq})$ decreases with respect to $J(x_{k|k}, U_{warm}(k))$,
\end{itemize}
then $U_{warm}(k)$ is replaced by $U(k)^{jq}$ and the algorithm continues backwards with respect to the prediction time $j$, in a similar manner. With this approach, at any point in time, if the  maximally allowed computational time is exceeded, a feasible, improved control sequence exists and it can be utilized as a suboptimal MPC solution.

By choosing to cover the control horizon backwards, we can reuse at each step $j$, the states $\phi(x_{k|k}, U_{warm}(k), i)$ for all $i\in\Zset_{[1, j]}$, which, by the feasibility of $U_{warm}(k)$, already satisfy the state and input constraints. This holds not only for the original $U_{warm}(k)$, but for any subsequent improvement of $U_{warm}(k)$. As such, also the stage costs up to the $j^{th}$ state can be recovered from previous computations.
The reusability of previously computed costs and states by navigating the control horizon in a backward manner resembles the working principle of DP. This suggests intuitively that the proposed cost improvement method could deliver good performance, which is supported by results obtained in non--trivial case studies, see Section~\ref{examples}. A formal analysis of convergence towards the DP solution, as the discrete--time $k$ increases, makes the object of future research.

When $k=0$, we can select an initial sequence $U_{warm}(0)$ as the solution of the optimal MPC problem. In this case, we can proceed with $k=1$. Alternatively, we can select a feasible sequence $U_{warm}(0)$, an ``oracle", by randomly selecting sequences of inputs in $\Uset$ until a feasible $U_{warm}(0)$ is found. In this case, if it is feasible to afford the computational time, we can proceed with Algorithm~\ref{alg} in an attempt of obtaining an improved sequence.


For $k\in\Zset_{\geq 1}$, to choose the input $U_{warm}(k)$ for Algorithm~\ref{alg}, one can use the receding horizon principle of MPC. As such, the input sequence $U_{warm}(k)=\{u_{k|k-1}, \ldots, u_{k+N-2|k-1}, u\}$ is a warm start at time $k$. If $\phi(x_{k-1|k-1}, U(k-1), N)\in\Xset_T$ and $\Xset_T$ is positively invariant for the system $x_{k+1}=f(x_k, k_f(x_k))$, one can select $u=k_f(x_{k+N-1|k-1}).$ In this case, $U_{warm}(k)$ is a feasible solution, and therefore a candidate warm start for every $k\in\Zset_{\geq 1}$. If there exists no terminal set $\Xset_T$ and no $k_f(\cdot)$, then one can select $u\in\Uset$ such that $U_{warm}(k)$ remains feasible, i.e., $\phi(x_{k-1|k-1}, U_{warm}(k), N)$ $\in\Xset$. In these circumstances, however, stability of the closed loop is not guaranteed. Such an example is illustrated in Section~\ref{wmr_section}.

\begin{remark}\label{halton}
Common sampling schemes are employed for sampling of the input space $\Uset$ at each iteration, among which we consider deterministic uniform sampling, which places each sample at equal distance from each other, to cover uniformly the space $\Uset$. Alternatively, ``true'' random samples can be selected, which are simpler to draw in higher dimensional spaces. Quasi--random low--discrepancy sequences, see \citep{chakrabarty2016}, may be used as well, considering the fact that they appear to be random for multiple purposes, such as Monte Carlo simulations. Such sampling methods, e.g., Sobol or Halton sequences, have been shown, see, e.g., \citep{de1994modelling}, to better cover the space than random sequences.
\end{remark}

\begin{remark}
Assume that $k_f$ is a locally stabilizing control law on $\Xset_T$, a sublevel set of $V_f$, which is positively invariant for the system $x_{k+1}=f(x_k, k_f(x_k))$ and $V_f$ and $L$ are, e.g., positive definite quadratic functions. Considering that the sequence $U(k)$ provided by Algorithm~\ref{alg} is a suboptimal solution solving Problem~\ref{prob}, then, by the arguments in Section~\ref{mpc}, Problem~\ref{prob} is recursively feasible and it ensures stability of system \eqref{eq:system} in closed loop with $u_k=u_{k|k}^w$.
\end{remark}

\begin{remark}
The working mechanism of Algorithm~\ref{alg} resembles the working principle of the rollout algorithm described in the survey paper \citep{bertsekas2005dynamic}. Therein, a rollout algorithm improves iteratively a base policy (here, the feasible control sequence $U_{warm}(k)$) to provide a suboptimal solution to an optimal control problem via approximative DP and suboptimal control. The working principle of the rollout algorithm with sampling based MPC, i.e., choosing at time $k$ the iterated solution of the previous time instance, $k-1$, as a warm start, and a sampling strategy, is not provided therein.
\end{remark}

\begin{remark}
Due to the sampling procedure having a discontinuous behaviour, the inputs might be varying more than it is safe for some applications. This problem might be alleviated by either smoothening the inputs via interpolation of the obtained input sequence with the initial sequence, or by penalizing $\Delta u_k=u_k-u_{k-1}$ via constraints or via the cost function, such that the input variability is limited to acceptable bounds.
\end{remark}

\subsection{Complexity analysis}
\label{complexity}

In order to analyze the complexity of Algorithm~\ref{alg}, the following assumptions are undertaken, for a given state $x$, input $u$ and input sequences $U$, $U_1$, $U_2$:
\begin{description}
\item[\textmd{1)}] The cost of evaluating $f(x, u)$ and performing a feasibility test $f(x, u)\in\Xset$ is $c_1$;
\item[\textmd{2)}] The cost of evaluating $J(x, U)$ is $c_2$;
\item[\textmd{3)}] The cost of comparing $J(x, U_1)<J(x, U_2)$ and changing $J_{sub}$ and $U_{warm}(k)$ if necessary, i.e., steps 11-13 in Algorithm~\ref{alg}, is negligible.
\item[\textmd{4)}] The current cost $J_{sub}$ is instantaneously available for comparison with each of the $n_j$ samples according to step 11 in Algorithm~\ref{alg} and each of the new sequences $U(k)^{jq}$ may modify $J_{sub}$ if the new cost $J_{new}$ is smaller than $J_{sub}$.
\end{description}
The complexity of Algorithm~\ref{alg} for a given $x_{k|k}$ is the following:
\begin{equation}\label{eq:compl1}
C=c_1\left(\sum_{j=0}^{N-1}(N-j)n_j\right)+c_2\left(\sum_{j=0}^{N-1}n_j\right).
\end{equation}
If we assume $n_j\leq \overline{n}$ for all $j\in\Zset_{[0, N-1]}$, then the complexity \eqref{eq:compl1} can be upper bounded by
\begin{equation}\label{eq:compl2}
C=\overline{n}c_1\frac{N(N+1)}{2}+c_2N\overline{n}.
\end{equation}
A possible reduction of the bound \eqref{eq:compl2} might be attained, considering the fact that all the states subsequent to a non--feasible state are no longer evaluated and checked for feasibility. This means that, in step 9 of Algorithm~\ref{alg}, if $\phi(x_{k|k}, U(k)^{jq}, i)\notin \Xset$ for a specific $i\in\Zset_{[j+1, N-1]}$, then $\phi(x_{k|k}, U(k)^{jq}, r)$ for all $r\in\Zset_{[i+1, N]}$, are no longer evaluated, in which case steps 2) and 3) are skipped all together.

Consider now that many threads are available, from multiple processors. Notice also that at each time in the horizon, all $n_j$ computations can be performed separately. In these conditions, assuming we have $\overline{n}$ threads available, then the complexity of Algorithm~\ref{alg} is upper bounded by
\begin{equation}\label{eq:nparallel}
C=c_1\frac{N(N+1)}{2}+c_2N.
\end{equation}
The complexity bound given in \eqref{eq:nparallel}, though quadratic in the prediction horizon, yields a reasonable complexity, considering that, in NMPC, a horizon $N=10$ is considered a reasonably large horizon.

In general, if we have $p\in\Zset_{[2, \overline{n})}$ processors, then the complexity of Algorithm~\ref{alg} is
\begin{equation}\label{eq:pparallel}
C=\ceil{\overline{n}/p}\left(c_1\frac{N(N+1)}{2}+c_2N\right),
\end{equation}
where the term $\ceil{\overline{n}/p}$ appears due to the fact that a thread can not engage in computations related to a subsequent time horizon until all the threads have finalized the computations related to the current time horizon $j$.

\section{Illustrative examples}
\label{examples}

The sampling--based suboptimal NMPC strategy proposed in Section~\ref{main_res} is illustrated on three nonlinear systems, to highlight various features of this method. All tests have been performed on a system with the following specifications: Intel Core i7-3770 CPU 3.4GHz, 16GB RAM, 64-bit OS.

\subsection{Cart--spring system}
The method developed in this paper will first be applied to a system incorporating an exponential nonlinearity, i.e.,  the model of a cart with mass $M$, which is moving on a plane, see \citep{raimondo2009min}. This cart is attached to a wall via a spring with elastic constant $k$ varying with the first state $k=k_0e^{-x_1}$, where $x_1$ stands for the displacement of the carriage from the equilibrium position. A damper acts as a resistor in the system, with damping $h_d$. The discretized nonlinear model of the cart and spring system is the following:
\begin{align}\label{eq:cart}
x(k+1)=\left[\begin{array}{c}
x_1(k+1) \\
x_2(k+1) \\
\end{array}\right]=f_1(x(k))+F_2u(k),
\end{align}
where
\begin{align}\label{eq:cart_details}
f_1(x(k))= & \left[\begin{array}{c}
x_1(k)+T_sx_2(k) \\
x_2(k)-T_s\frac{\rho_0}{M}e^{-x_1}x_1(k)-T_s\frac{h_d}{M}x_2(k) \\
\end{array}\right],  \nonumber \\
F_2= & \left[\begin{array}{cc}
0 & \frac{T_s}{M}
\end{array}\right]^T,
\end{align}
where $x_2$ is the velocity of the cart and $u$ is an external force which acts as an input to the system. The parameter values are $T_s=0.4s$, $\rho_0=0.33$, $M=1$, $h_d=1.1$.

The MPC controller  has to steer the cart to the origin from a non--zero initial state, while satisfying the input and state constraints, which are
\begin{equation}\label{eq:cart_constr}
|u|\leq 4.5, |x_1|\leq 2.65,
\end{equation}
and reducing the cost \eqref{eq:cost}, where the stage cost and terminal cost are quadratic functions, i.e., $L(x,u)=x^TQx+u^TRu$ and $V_f(x)=x^TPx$.
Choose the following parameters for the MPC problem:
$$M=4, P=\left[\begin{array}{cc}
7.0814 & 3.3708 \\
3.3708 & 4.2998 \\
\end{array}\right], Q=diag(1, 1), R=1.$$
In \citep{raimondo2009min} it is shown that the control law
$$u=k_f(x)=-\left[\begin{array}{cc}
0.8783 & 1.1204
\end{array}\right]f_1(x),$$
is locally stabilizing in the set
$$\Xset_T=\{x | V_f(x)\leq 4.7\},$$

\begin{figure}[!htpb]
      \centering
      \includegraphics[width=1\columnwidth]{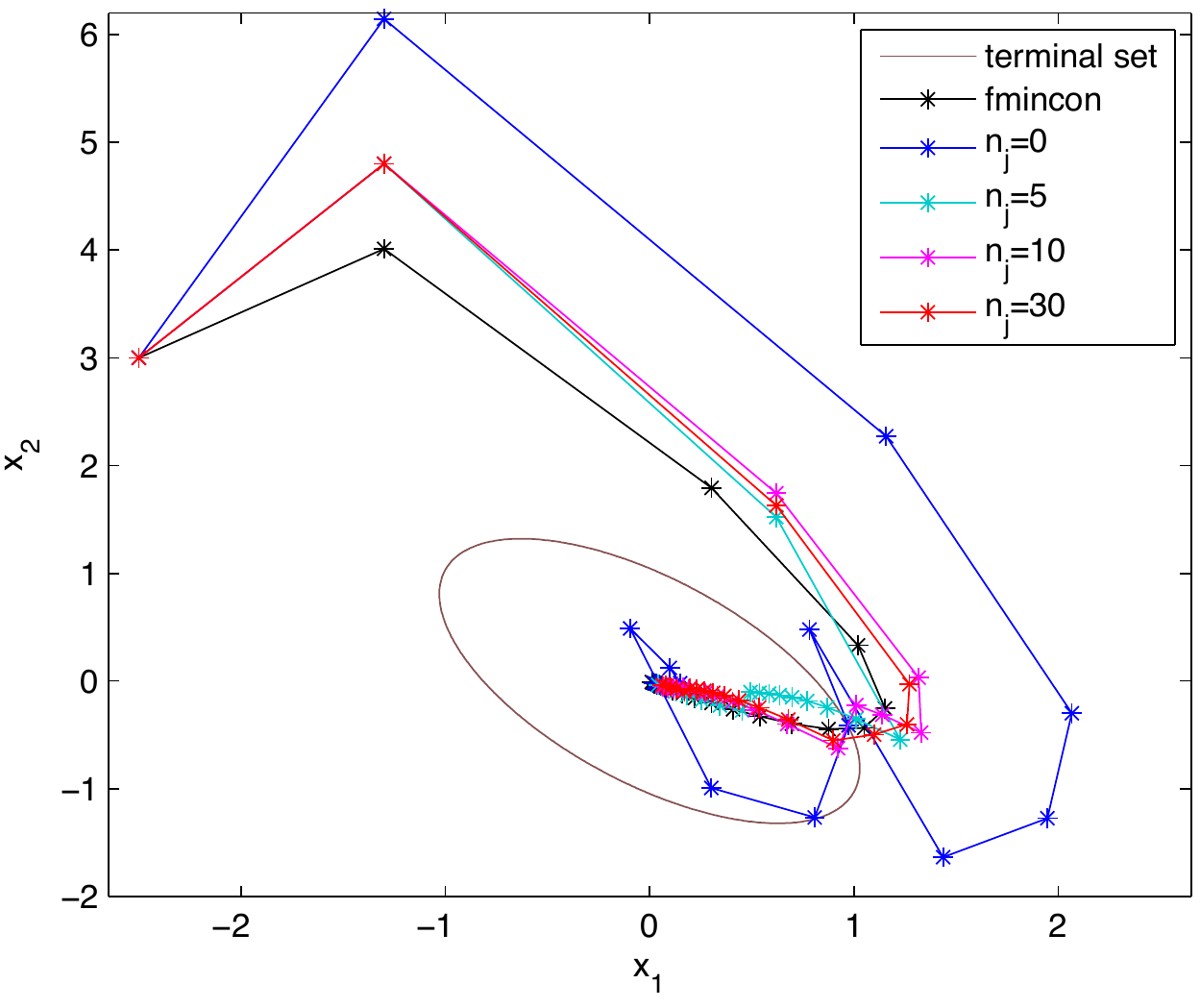}
      \caption{State trajectory.}
      \label{state1}
\end{figure}

\begin{figure}[!htpb]
      \centering
      \includegraphics[width=1\columnwidth]{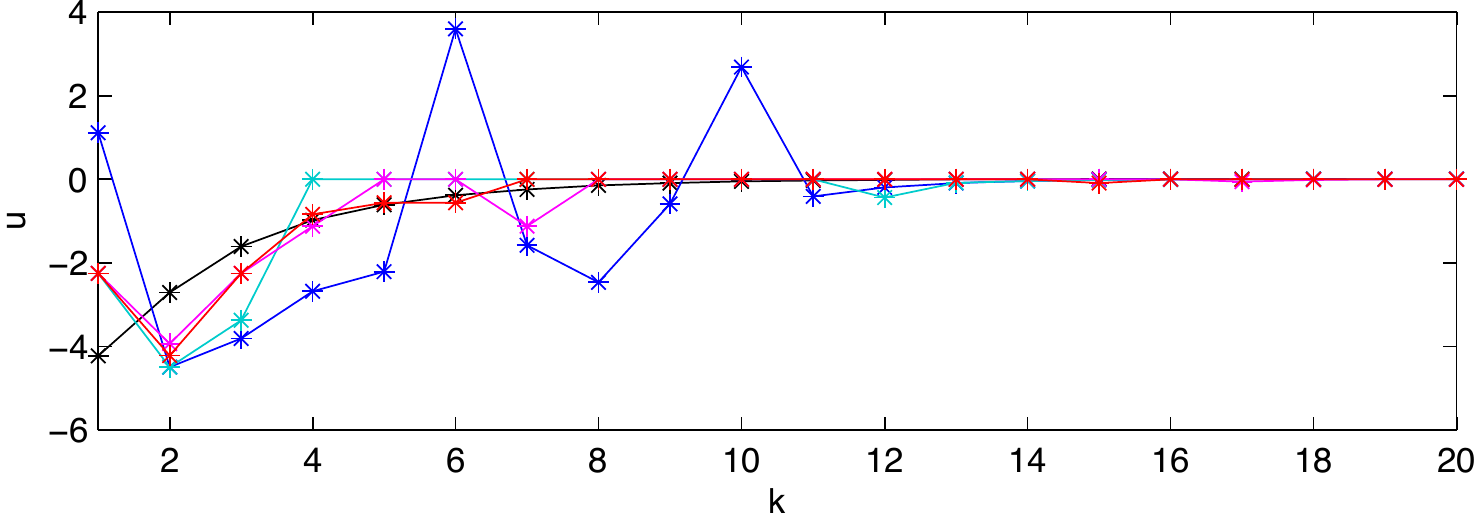}
      \caption{Inputs applied to the system.}
      \label{inputs1}
\end{figure}

which is a terminal set where the conditions for stability of MPC mentioned in Section~\ref{mpc} are satisfied.

Algorithm~\ref{alg} is applied for the MPC control of system \eqref{eq:cart}. We compare the results of this method with the results provided by \verb"fmincon" in Matlab, even though, as it will be seen later, the optimization tool does not always provide the optimum, due to local minima. The scalability of the algorithm is tested by varying both the number of samples $\overline{n}$ and the control horizon $N$. The tests illustrated in this paper are performed on a feasible initial state $x_{0|0}=[-2.5, 3]$. The choice of the initial condition does not influence greatly the results, which are similar for other feasible initial states.

\begin{figure}[!htpb]
      \centering
      \includegraphics[width=1\columnwidth]{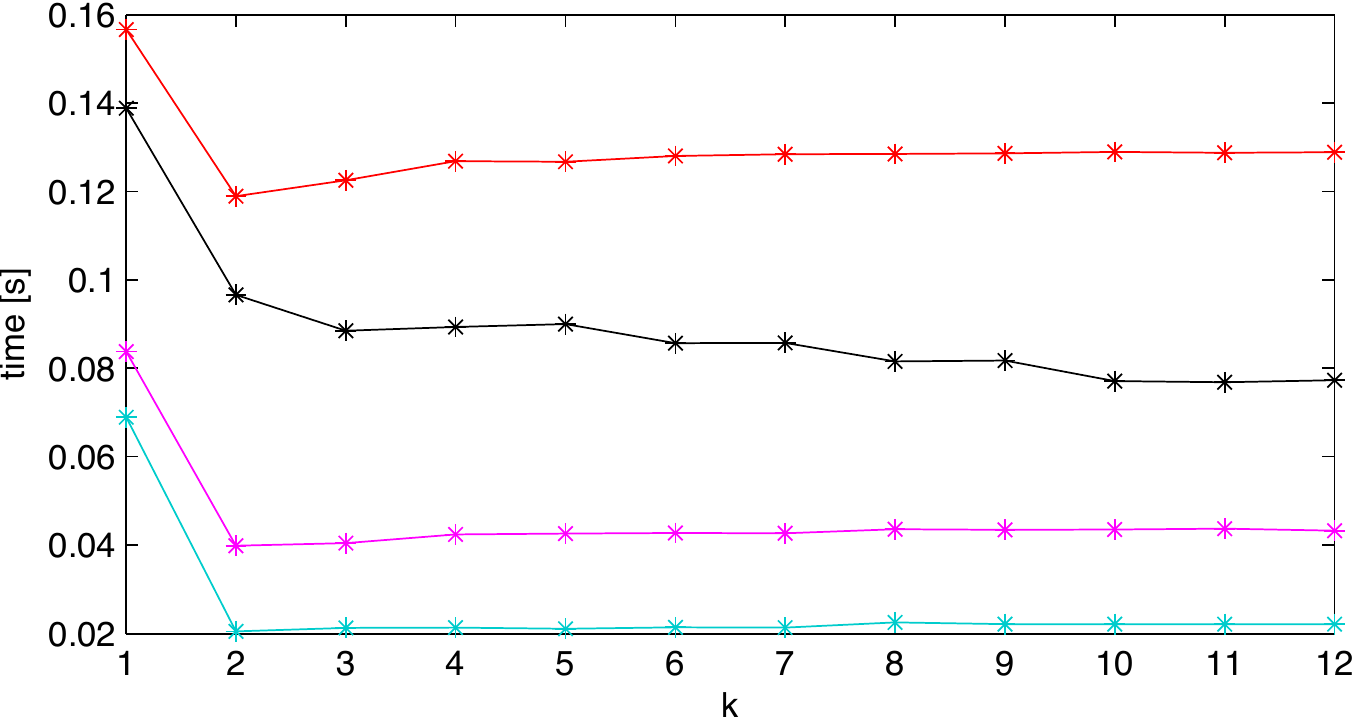}
      \caption{Computation time.}
      \label{times1}
\end{figure}

\begin{figure}[!htpb]
      \centering
      \includegraphics[width=1\columnwidth]{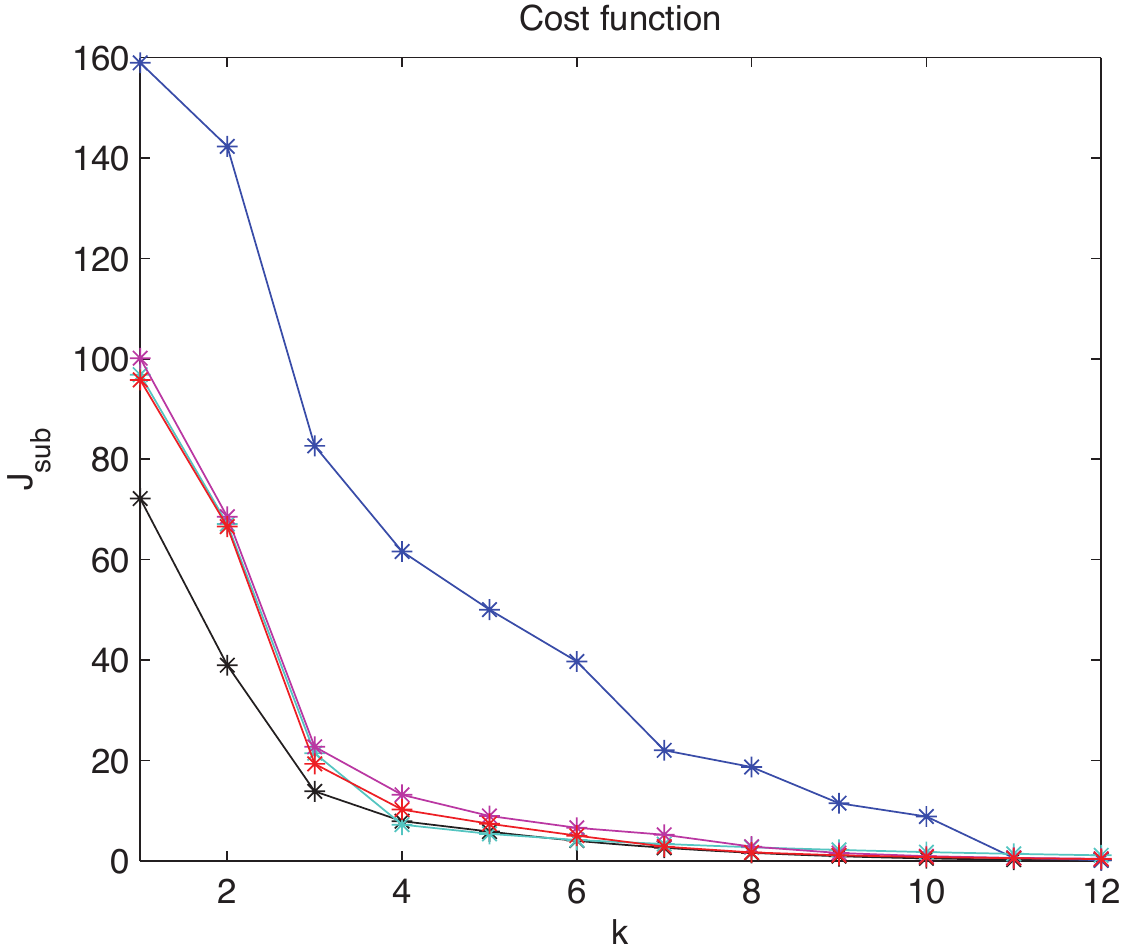}
      \caption{The cost $J_{sub}$.}
      \label{costs1}
\end{figure}

For the first experiment, fix $N=10$. An initial $U_{warm}(0)$ is provided by a random ``oracle". Consider $n_j=\overline{n}$, for all $j\in\Zset_{[0, N-1]}$, taking various values in the set $\{0, 5, 10, 30\}$. For $\overline{n}=0$, $U_{warm}(0)$ is propagated though iterations without any intervention or change from the sampling mechanism. This serves as a reference, to notice the improvements brought in by Algorithm~\ref{alg}. The results are illustrated in Fig.~\ref{state1}--\ref{costs1}, where the legend from Fig.~\ref{state1} holds until Fig.~\ref{costs1}. Iterations are considered from $k=1$ until $k=20$. Though the different sampling options proposed in Remark~\ref{halton} provide in this example similar outcomes, Halton points have been used here for illustration. This choice is motivated by the practical feature that, adding extra points only when they are required does not have impact on the coverage of the set $\Uset$.

\begin{figure}[!htpb]
      \centering
      \includegraphics[width=1\columnwidth]{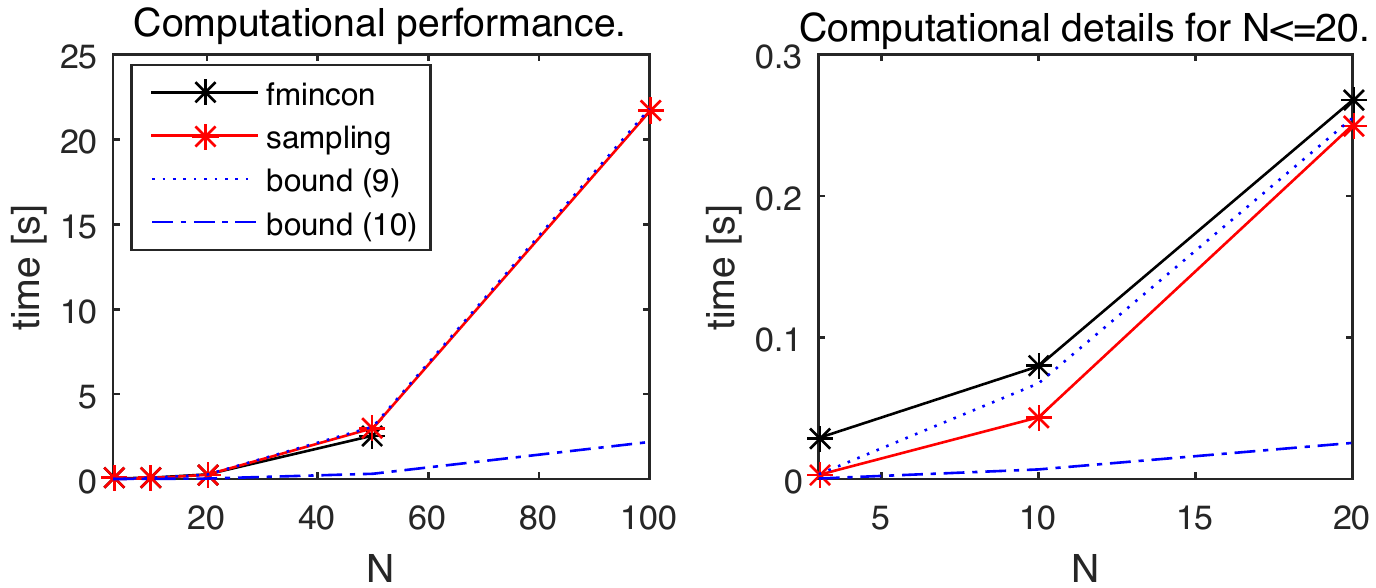}
      \caption{Computational performance versus proposed bounds.}
      \label{horizon1}
\end{figure}

In Fig.~\ref{state1} and Fig.~\ref{inputs1} the state trajectory and the inputs $u(k)$ applied to the system are illustrated. The constraint specifications \eqref{eq:cart_constr} are satisfied for all cases, at all times. Notice the immediate smoothening of the trajectories and input sequence even for a small $\overline{n}$. In Fig.~\ref{times1}, the computational time, without parallelization, is illustrated. At $k=1$, the computational cost of finding an oracle is included. Fig.~\ref{costs1} illustrates the values of $J_{sub}$ for each iteration. Notice, overall, that even for a small number of samples, the performance of the closed loop system is significantly improved and the computational time is promising, even for a non--parallel implementation. Also, as the iteration $k$ advances, the initially modest performance increases significantly, due to the continuous improvement of $U_{warm}$ and the receding horizon principle. Interestingly, even for $\overline{n}=5$, at iteration $k=4$, the cost $J_{sub}(4)$ is smaller than the cost
computed via \verb"fmincon", which, due to local minima, provided a feasible but not optimal solution.

For the second experiment we aim to test the computational complexity of Algorithm~\ref{alg} in terms of the control horizon. The implementation used here does not use parallelization. Fix $\overline{n}=10$ and $N$ takes values in the set $\{3, 10, 20, 50, 100\}$. The results are illustrated in Fig.~\ref{horizon1}. With red, the computational complexity for Algorithm~\ref{alg} is depicted, and it is always smaller than the bound \eqref{eq:compl2}, which is followed closely. It is expected that on dedicated devices, the complexity of Algorithm~\ref{alg} is smaller, due to the processors not running in parallel threads related to other system applications. With parallelization, further reduction in complexity is expected, as described in Section~\ref{complexity} and illustrated in Fig.~\ref{horizon1}. Notice that, for smaller horizons, the complexity of Algorithm~\ref{alg} is smaller than the complexity of \verb"fmincon", even without parallelization. For $N=100$, \verb"fmincon" provides solutions which are not feasible, while the proposed Algorithm~\ref{alg} still terminates in 21.6 seconds.

\subsection{Buck--Boost power converter}
Next, the bilinear model of a Buck--Boost power converter is considered, as in \citep{spinu2011explicit}:
\begin{align}\label{eq:buck}
x(k+1)=Ax(k)+Bu(k)+\left[\begin{array}{c}
x(k)^TC_1 \\
x(k)^TC_2 \\
\end{array}\right]u(k),
\end{align}
where $x:=\left[\begin{array}{cc}
v_C & i_L
\end{array}\right]^T\in\Xset\subset\Rset^2$ is the state vector consisting of the voltage
across the output capacitor
and the current
through the filter inductor.
The input $u:=\left[\begin{array}{cc}
d_1 & d_2
\end{array}\right]^T\in\Uset\subset\Rset^2$ stands for the duty--cycle ratio of the control signal applied to the switching node.
The parameters are
$$A=\left(I_2+T_s \left[\begin{array}{cc}
-\frac{1}{R_HC} & 0\\
0 & -\frac{R_L}{L} \\
\end{array}\right]\right), B=\left[\begin{array}{cc}
0 & 0\\
\frac{v_s}{L} & 0 \\
\end{array}\right]T_s,$$
$$C_1= \left[\begin{array}{cc}
0 & 0\\
0 & \frac{1}{C} \\
\end{array}\right]T_s, C_2=\left[\begin{array}{cc}
0 & -\frac{1}{L}\\
0 & 0 \\
\end{array}\right]T_s,$$
with the values $R_L =
0.2\Omega$, $C = 22\mu F$, $L = 220\mu H$, $T_s = 10\mu s$.

The aim of the control loop is to stabilize the system to the equilibrium point $x_{e}:=\left[\begin{array}{cc}
20 & 0.5
\end{array}\right]^T$, $u_{e}:=\left[\begin{array}{cc}
0.81 & 0.4
\end{array}\right]^T$, under the constraints $i_L\in\Rset_{[0, 3]}$, $v_C\in\Rset_{[-0.1, 22.5]}$, $u\in\Rset^2_{[0, 1]}$. The terminal controller

\begin{figure}[!htpb]
      \centering
      \includegraphics[width=1\columnwidth]{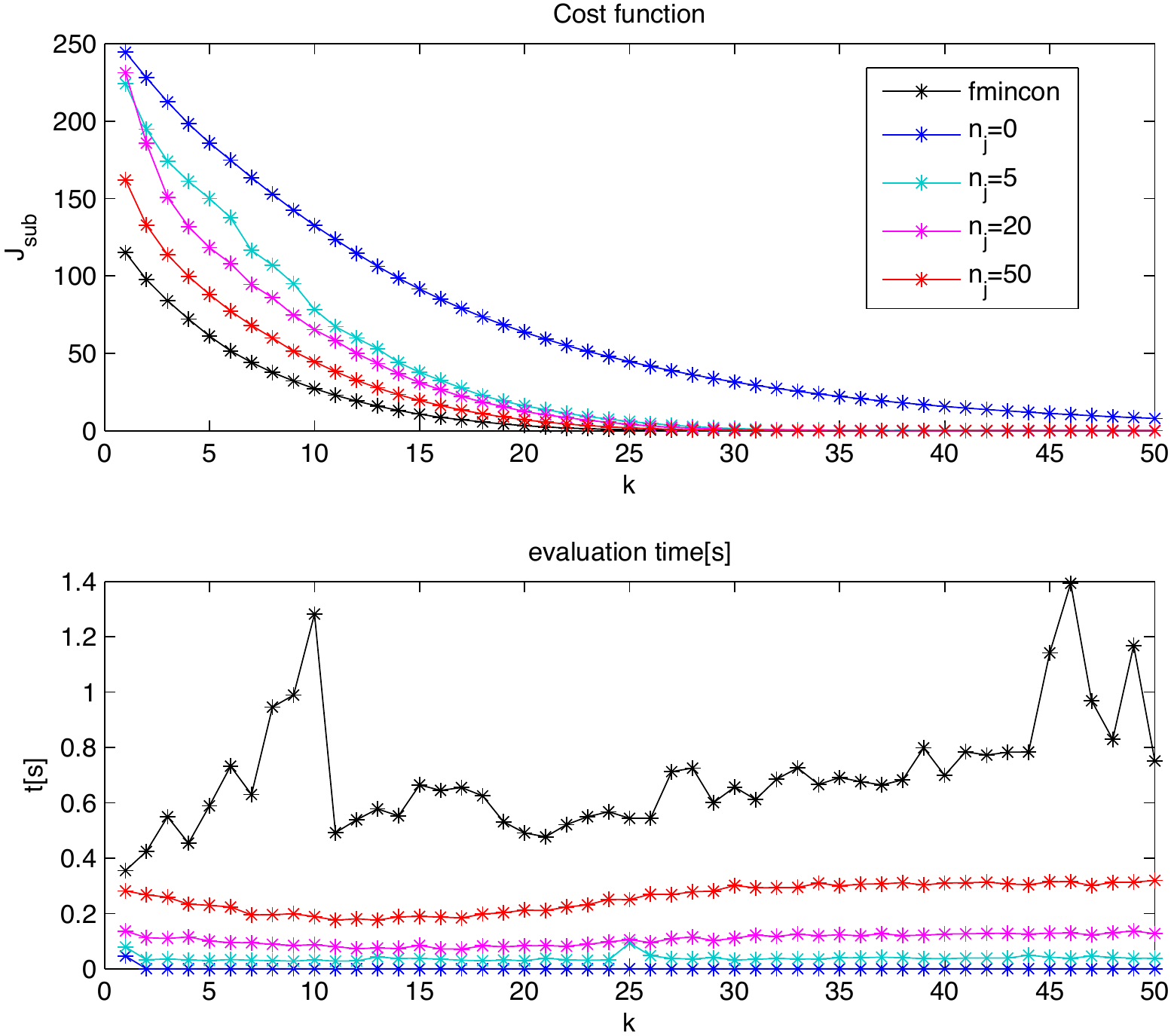}
      \caption{Performance of Algorithm~\ref{alg} on the power converter.}
      \label{horizon2}
\end{figure}

$$u=u_e+K(x-x_e), K=\left[\begin{array}{cc}
   -0.0014  & -0.3246 \\
    0.0001  & -0.0055 \\
\end{array}\right],$$
stabilizing the system in the terminal set $\Xset_T$ given in \citep[set $\Pset$ in Section IV.C]{spinu2011explicit}, and the quadratic cost with the matrices
$$Q=diag(1, 2), R=diag(1, 1), P=\left[\begin{array}{cc}
    46.6617  & 42.8039 \\
   42.8039  & 69.4392 \\
\end{array}\right], $$
satisfies all the conditions for stability formulated in Section~\ref{mpc}.

Similarly to the previous example, we apply Algorithm~\ref{alg} for the MPC control of system \eqref{eq:buck}. Fix the initial state $x_{0|0}=
\left[\begin{array}{cc}
   1  & 2 \\
\end{array}\right]^T+x_e$, which is outside of the terminal set $\Xset_T$, and $N=10$. $U_{warm}(0)$ is given by an oracle and we use random sampling of $\Uset$. See in Fig.~\ref{horizon2} the effect of varying $\overline{n}$ on the cost function $J_{sub}$ and the time necessary, per iteration, to compute the corresponding control law. Notice the effect of the unknown termination time on the evaluation time for the optimization performed through \verb"fmincon" and the relatively equal computational time of Algorithm~\ref{alg} over the iterations $k$. All the constraints were satisfied for all the presented situations, and it is expected that, through implementation on dedicated multi--thread systems, the computational time for the control law decreases to the extent of fitting the tight sampling period of the power converter.

\subsection{Wheeled mobile robot}
\label{wmr_section}
The last example illustrates the methodology developed in this paper for an obstacle avoidance task by a nonholonomic system with trigonometric nonlinearities, due to kinematics, i.e., a model of the wheeled mobile robot (WMR), as described in \citep{kuhne2005point}:
\begin{align}\label{eq:wmr}
x(k+1)=\left[\begin{array}{c}
x_1(k)+u_1(k)\cos x_3(k) T_s\\
x_2(k)+u_1(k)\sin x_3(k) T_s \\
x_3(k)+u_2(k)T_s \\
\end{array}\right].
\end{align}
In \eqref{eq:wmr}, the state $x\in\Rset^3$ describes the position and the orientation of the robot with respect to a global inertial frame $\{O, X, Y\}$, and the input $u\in\Rset^2$ gives the linear and angular velocity, respectively. The parameter $T_s=0.1s$ is the discretization period of system \eqref{eq:wmr}.

The MPC strategy aims at driving the WMR from an initial state $x_{0|0}=
\left[\begin{array}{ccc}
0 & 6 & 0\\
\end{array}\right]^T$ to the origin of the inertial frame, i.e., $x_{g}=\left[\begin{array}{ccc}
0 & 0 & 0\\
\end{array}\right]^T$, while satisfying the input  constraints $u_1\in\Rset_{[-0.47, 0.47]}$, $u_2\in\Rset_{[-3.77, 3.77]}$. A quadratic cost function of the form
\begin{align}
J(x_{k|k}, U(k))= & x_{k+N|k}^TPx_{k+N|k}+\sum_{j=1}^{N-1}x_{k+j|k}^TQ(j)x_{k+j|k}+ \nonumber\\
& +\sum_{j=0}^{N-1}u_{k+j|k}^TRu_{k+j|k} \nonumber
\end{align}
is considered, with the parameters: $Q(j)=2^{j-1}Q$, $P=50Q(N)$, $Q=diag(1, 1, 0.5)$, $R=diag(0.1, 0.1)$, $N=5$.

\begin{figure}[!htpb]
      \centering
      \includegraphics[width=1\columnwidth]{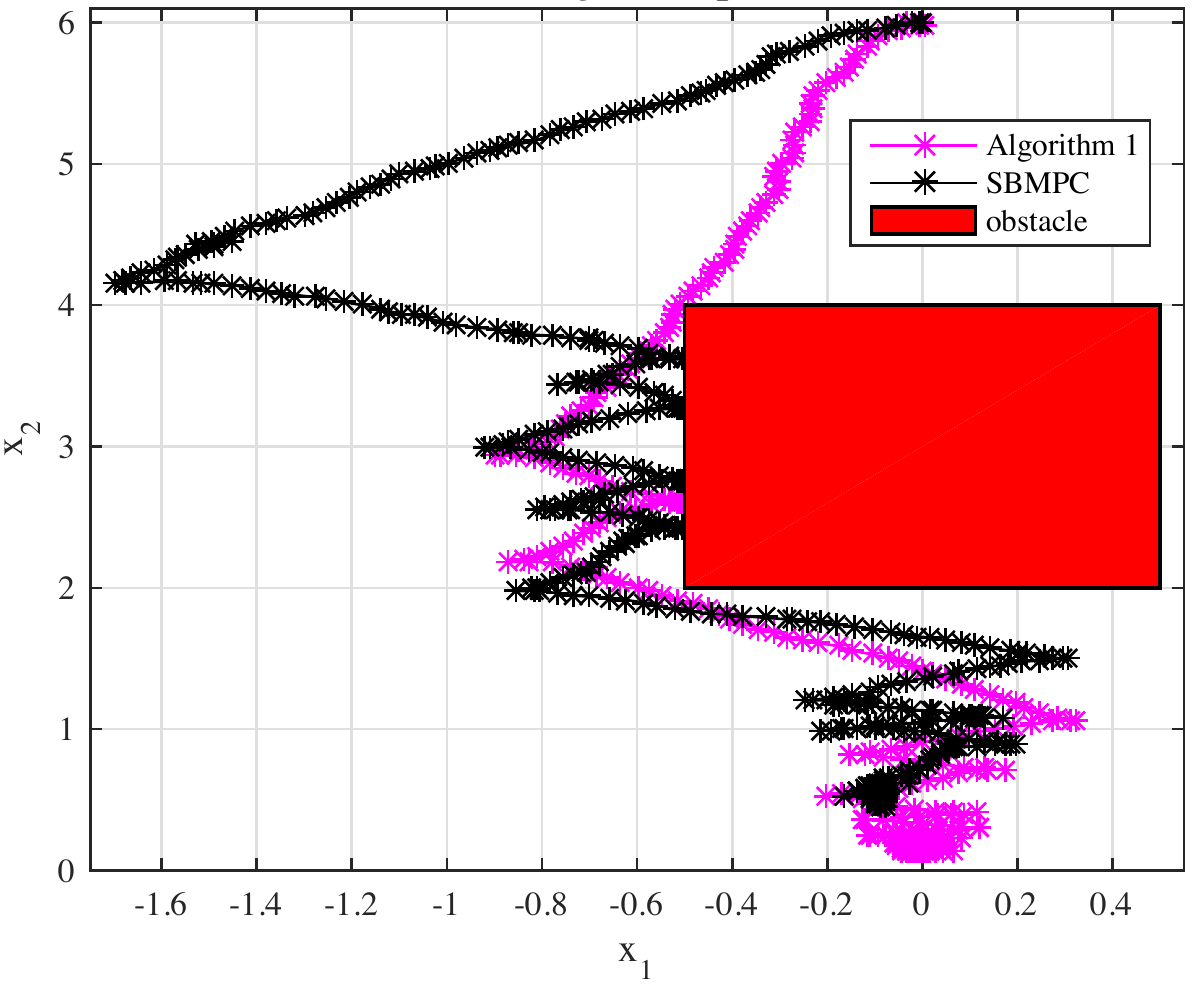}
      \caption{Trajectories of WMR.}
      \label{robot1}
\end{figure}
\begin{figure}[!htpb]
      \centering
      \includegraphics[width=1\columnwidth]{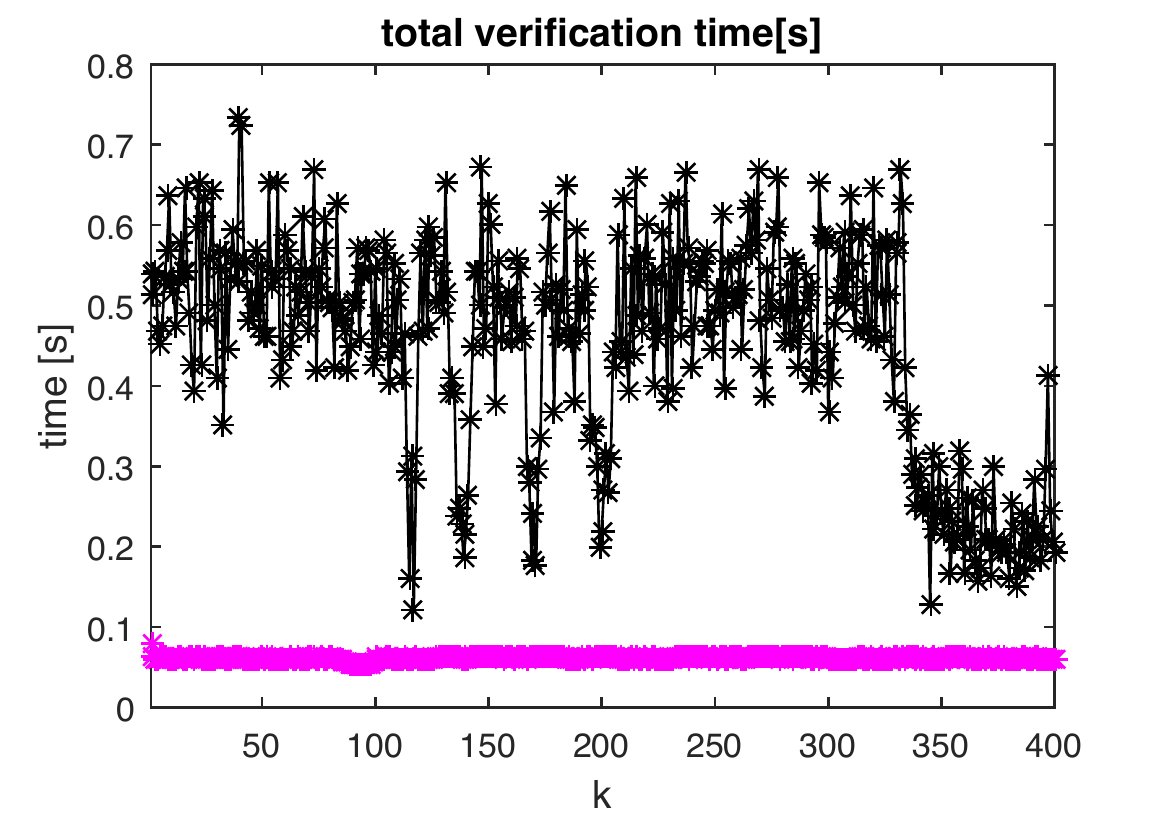}
      \caption{Time performance of WMR.}
      \label{robot2}
\end{figure}

In this example we illustrate Algorithm~\ref{alg} with the above parameters. Consider $U_{warm}(0)$ generated by an ``oracle", and sampling of $\Uset$ based on random sequences. The result for $\overline{n}=30$ is illustrated in Fig~\ref{robot1}. \verb"fmincon" could not be applied due to feasibility issues related to non--convexity caused by the presence of the obstacle. Notice the avoidance of the obstacle of the WMR under the control law generated by Algorithm~\ref{alg}. Due to the sampling mechanism, however, the inputs are not smooth, which could be alleviated, e.g., by an interpolation mechanism. For comparison we illustrate also the result using SBMPC \citep{dunlap2010nonlinear}, with the same horizon $N$ and $\overline{n}=30$. Notice, in the state trajectory plot in Fig~\ref{robot1} the fact that, after 400 iterations, the WMR did not reach $x_g$ yet, 
and the input values are not yet 0, which means that the sampling mechanism did not consider that the WMR is close to the goal. In this case, it is recommended to sample more densely around 0 in the set $\Uset$, rather than uniformly covering $\Uset$ with samples.

Notice in Fig~\ref{robot2} that, due to the building of a tree and the search in a tree for the best path, SBMPC is more computationally demanding. The complexity of Algorithm~\ref{alg} with the given $\overline{n}$ fits the sampling period $T_s=0.1s$ of the WMR, which makes the method applicable for real--time control of this system.


\section{Conclusion}
\label{conclusions}

In this paper, an algorithm based on sampling of the input space at each time in the horizon was proposed, which iteratively improves in terms of cost an initially feasible control sequence. This suboptimal NMPC strategy provides a promising computational complexity even for large control horizons, with good perspectives for parallel implementation. Future work aims at investigating the convergence of the proposed algorithm to the optimal solution, the scalability in terms of system dimension and the effect of parallelization on computational time for real--life applications.


\bibliography{Ruxandra}             








\end{document}